\documentclass[12pt]{article}

\newcommand{\pmSig}{\,^{\pm}\!\Sigma}
\newcommand{\pSig}{\,^+\!\Sigma}
\newcommand{\mSig}{\,^-\!\Sigma}
\newcommand{\pbpi}{\,^+\!\mbox{\boldmath$\pi$}}
\newcommand{\pbv}{\,^+\!\mbox{\boldmath$v$}}
\newcommand{\mbv}{\,^-\!\mbox{\boldmath$v$}}
\newcommand{\pbtau}{\,^+\!\mbox{\boldmath$\tau$}}
\newcommand{\mbtau}{\,^-\!\mbox{\boldmath$\tau$}}
\newcommand{\mbpi}{\,^-\!\mbox{\boldmath$\pi$}}
\newcommand{\bpi}{\mbox{\boldmath$\pi$}}
\newcommand{\bphi}{\mbox{\boldmath$\phi$}}
\newcommand{\btau}{\mbox{\boldmath$\tau$}}

\renewcommand{\d}{{\rm d}}

\newcommand{\pv}{\,^+\!v}
\newcommand{\mv}{\,^-\!v}
\newcommand{\pmbSig}{\,^{\pm}\!\mbox{\boldmath$\Sigma$}}

\newcommand{\pmbphi}{\,^{\pm}\!\mbox{\boldmath$\phi$}}
\newcommand{\pmR}{\,^{\pm}\!R}
\newcommand{\pmphi}{\,^{\pm}\!\phi}
\newcommand{\pmbn}{\,^{\pm}\!\mbox{\boldmath$n$}}
\newcommand{\pmu}{\,^+\!\mu}
\newcommand{\mmu}{\,^-\!\mu}
\newcommand{\pmmu}{\,^{\pm}\!\mu}
\newcommand{\pmtau}{\,^{\pm}\!\tau}
\newcommand{\pmpi}{\,^{\pm}\!\pi}
\newcommand{\pmbpi}{\,^{\pm}\!\mbox{\boldmath$\pi$}}
\newcommand{\bn}{\mbox{\boldmath$n$}}
\newcommand{\pmcalN}{\,^{\pm}\!{\cal N}}
\newcommand{\pcalN}{\,^+\!{\cal N}}
\newcommand{\mcalN}{\,^-\!{\cal N}}

\begin{document}

\title{Path integral in area tensor Regge calculus and complex connections}
\author{V.M. Khatsymovsky \\
 {\em Budker Institute of Nuclear Physics} \\ {\em
 Novosibirsk,
 630090,
 Russia}
\\ {\em E-mail address: khatsym@inp.nsk.su}}
\date{}
\maketitle
\begin{abstract}
Euclidean quantum measure in Regge calculus with independent area tensors is
considered using example of the Regge manifold of a simple structure. We go over to
integrations along certain contours in the hyperplane of complex connection variables.
Discrete connection and curvature on classical solutions of the equations of motion
are not, strictly speaking, genuine connection and curvature, but more general
quantities and, therefore, these do not appear as arguments of a function to be
averaged, but are the integration (dummy) variables. We argue that upon integrating
out the latter the resulting measure can be well-defined on physical hypersurface (for
the area tensors corresponding to certain edge vectors, i.e. to certain metric) as
positive and having exponential cutoff at large areas on condition that we confine
ourselves to configurations which do not pass through degenerate metrics.
\end{abstract}
\newpage
The functional integral approach remains the most efficient method of quantization of
such the discrete version of general relativity as Regge calculus \cite{Reg} (RC).
General conception of this approach, definitions and theorems were discussed in
\cite{Fro}. Functional integral is a basic tool in investigation of the properties of
quantum RC \cite{HamWil1,HamWil2} (one of the consequences of this study being
occurrence of the phase transitions in simplicial quantum gravity). For a review of RC
and alternative discrete gravity approaches see, e. g., \cite{RegWil}.

Functional integral in the continuum theory is usually defined as Feynman path
integral based on the canonical (Dirac) quantization prescription. Since continuous
time coordinate is absent in RC, now we cannot develop Hamiltonian formalism and
canonical quantization. So we are free to choose another physical requirements to
hopefully fix completely discrete functional integral measure, and different sets of
these requirements do not necessarily result in the same measure. Standard choice of
the measure was considered in \cite{HamWil2}. The choice of interest for us (a
posteriori motivated by the hopes to get absolutely convergent functional integral) is
to require that some limiting continuous time form of the functional integral of
interest be exactly Feynman path integral in the limiting continuous time (so-called
(3+1)) RC.

The present paper continues our work on studying the functional integral defined in
such the way. Let us briefly summarize main points in the solution to this problem. A
coordinate is made continuous by shrinking sizes of all the simplices along this
coordinate to those infinitely close to zero. Then one regards this coordinate as a
formal time and tries to develop Hamiltonian formalism and canonical quantization and
represent the result in the form of the Feynman path integral measure.

Next it is natural to ask whether a measure in the original completely discrete RC
exists such that one could define the formal limit of this measure when one of the
coordinates is made continuous and the limiting measure would coincide with the above
Feynman path (canonical quantization) measure with this continuous coordinate playing
the role of time. Equivalence of the different coordinates means that this situation
should take place irrespectively of what coordinate is made continuous.

A difficulty with RC in the continuous time limit is that the description of the
infinitely flattened in some direction simplex purely in terms of the lengths is
singular. The idea is to use description in terms of the variables of the types of
both lengths and angles. This might be achieved in the Regge analog of the
Hilbert-Palatini form of the Einstein action. The discrete analogs of the tetrad and
connection, edge vectors and finite rotation matrices, were first considered in
\cite{Fro}. Approximate representation (for small deficit angles) of the Regge action
in terms of edge vectors and independent finite rotation matrices has been proposed in
the paper \cite{CasDadMag}. Simultaneously, we have proposed in our paper \cite{Kha0}
representations (ordinary and selfdual ones) of the Regge action for arbitrary deficit
angles. The representation of this type results in the exact Regge action if rotation
matrices are excluded via equations of motion. Rotation matrices are just the desired
angle type variables which allow us to formulate continuous time (3+1) RC in a
nonsingular way. After that the above strategy can be implemented. In \cite{Kha3} we
write out canonical form of RC in the considered variables.

Next we can try to solve the problem of finding measure in the full discrete RC which
has the desired continuous time limit corresponding to the canonical quantization
irrespectively of what coordinate is taken as a time and made continuous. Although
this last condition is rather restrictive, the problem has solution in 3 dimensions
\cite{Kha2}. In 4 dimensions solution can be found for a certain version of the
so-called "area RC" where areas are treated as independent variables
\cite{BarRocWil,RegWil}. Since the number of areas is larger than the number of
lengths, this means that the lengths of the same edge defined in the different
simplices are in general different, i.e. ambiguous. The configuration superspace of
the area RC contains the hypersurface corresponding to the ordinary RC; at the same
time it is exactly soluble just as 3D model. Appropriate version in our case is "area
tensor RC" with independent {\it area tensors} (i.e., in particular, in general there
are no edge vectors corresponding to them). Just the corresponding superspace
(extended as compared to that of ordinary RC) is that space on which the measure under
above restrictive conditions can be found \cite{Kha}.

Remarkable feature simplifying solution of the above problem of finding full discrete
measure for the 3D RC and for the 4D area tensor RC in the tetrad-connection variables
is commutativity of the constraints (arising in the Hamiltonian formalism, i.e. in the
continuous time limit). These constraints as well as canonical quantization itself are
analogous to their completely continuum counterparts. The commuting constraints for
the 3D discrete gravity were found in \cite{Wael} for general system (not a'priori
restricted to be RC). Analogous first class system of constraints arises in the area
tensor RC in the above mentioned derivation of the measure \cite{Kha}.

Finally, we need to reduce the quantum measure in the extended superspace to the RC
hypersurface. The idea is to consider area tensor RC system and ordinary RC system as
particular case of the simplicial complex with discontinuous metrics. The point is
that the piecewise flat manifold possesses metric whose normal component undergoes
discontinuity when passing across any 3D face, but the tangential components remain
unchanged. Now we go further and consider system where tangential components of metric
are also discontinuous. It is the system with independent simplices which do not
necessarily fit each other on their common faces. In the superspace of all the
simplicial discontinuous metrics RC corresponds to the hypersurface singled out by
conditions of the tangential metric continuity on the faces. The problem is to reduce
the above constructed quantum measure in "area tensor RC" to this hypersurface. For
that some $\delta$-function-like factor is introduced in the measure which fixes
equality of tangential metric across any face. This factor is found in our paper
\cite{Kha1} by using the principle of "minimum of the lattice artefacts". Namely, we
require that the factor should not depend on the form and size of any face across
which metrics are compared, only on the hyperplane in which the given face is placed.
We show that such the factor preserving equivalence of the different simplices exists
and is unique. Consequences of the viewpoint on area RC (now not tensor one) as a
system with discontinuous metrics were also discussed in \cite{WaiWil}.

The result of \cite{Kha} for the Euclidean measure being applied to arbitrary function
on the set of area tensors $\pi^{ab}_{\sigma^2}$ and connection matrices
$\Omega^{ab}_{\sigma^3}$ reads
\begin{eqnarray}
\label{VEV}                                                                         
<\Psi (\{\pi\},\{\Omega\})> & = & \int{\Psi (-i\{\pi\}, \{\Omega\})\exp{\left (-\!
\sum_{t{\rm -like}\, \sigma^2}{\tau _{\sigma^2}\circ R_{\sigma^2}(\Omega)}\right
)}}\nonumber\\
 & & \hspace{-20mm}\times \exp{\left (i
\!\sum_{{\rm not}\, t{\rm -like}\, \sigma^2} {\pi_{\sigma^2}\circ
R_{\sigma^2}(\Omega)}\right )}\prod_{\stackrel{\stackrel{\rm
 not}{t-{\rm like}}}{\sigma^2}}{\rm d}^6
\pi_{\sigma^2}\prod_{\sigma^3}{{\cal D}\Omega_{\sigma^3}} \nonumber\\ & \equiv &
\int{\Psi (-i\{\pi\},\{\Omega\}){\rm d} \mu_{\rm area}(-i\{\pi\},\{\Omega\})}.
\end{eqnarray}

\noindent Here $A\circ B$ $\stackrel{\rm def}{=}$ ${1\over 2}A^{ab}B_{ab}$, ${\cal
D}\Omega_{\sigma^3}$ is the Haar measure on the group SO(4) of connection matrices
$\Omega_{\sigma^3}$, the curvature $R_{\sigma^2}(\Omega)$ is path-ordered product of
connections $\Omega_{\sigma^3}$ along the path enclosing the triangle
$\Omega_{\sigma^2}$. Here rotation of the integration contours used to define integral
is performed via substitution of the integration variables $\pi_{\sigma^2}$ $\to$
$-i\pi_{\sigma^2}$. The $\pi_{\sigma^2}$ and $\tau_{\sigma^2}$ are the area tensors
for the "not t-like" and "t-like" triangles, respectively, while arbitrary tensor
($\pi_{\sigma^2}$ or $\tau_{\sigma^2}$) will be denoted $v_{\sigma^2}$. The
definitions of these triangle are intuitively understandable, the notion "t-like"
should refer to a triangle (or, more generally, to a simplex of arbitrary
dimensionality) with one of the edges connecting the pair of corresponding points in
the two analogous in structure neighbouring {\it leaves of the foliation} (3D Regge
structures themselves) along one of the coordinates named time t. The usual definition
"timelike" is reserved to be attribute of the local frame indices $a$, $b$ in
$v^{ab}_{\sigma^2}$ while the above "t-like" refers to $\sigma^2$, the Regge analog of
the pair of the world indices.

The formula (\ref{VEV}) looks as the usual field-theoretical path integral expression
with exception of the following three points. First, occurrence of the SO(4) Haar
measure ${\cal D}\Omega_{\sigma^3}$ instead of the Lebesgue measure on the
infinitesimal connections in the continuum case. This is connected with the specific
form of the kinetic term $\pi_{\sigma^2}\circ
{\Omega}^{\dag}_{\sigma^2}\dot{\Omega}_{\sigma^2}$ obtained when we are going over to
the continuous time limit (here $\Omega_{\sigma^2}$ serves to parameterize limiting
form of $\Omega_{\sigma^3}$ with $\sigma^3$ filling up the infinitesimal t-like
3-prism with the base $\sigma^2$). Canonical quantization of the theory with such the
kinetic term just gives ${\cal D}\Omega_{\sigma^2}$. Tracing back to the original full
discrete theory, this means occurrence of ${\cal D}\Omega_{\sigma^3}$ there too.

Second, there are the terms $\pi\circ R$ in the exponential instead of the Regge
action which in the exact connection representation would be the sum of the terms with
'arcsin' of the type $|\pi |\arcsin{(\pi\circ R/|\pi |)}$ \cite{Kha0}. The reason is
that the exponential should be the sum of the constraints times Lagrange multipliers
in order to fit the canonical quantisation prescription in the continuous time limit
whatever coordinate is taken as a time. These constraints just do not contain 'arcsin'
(in empty spacetime; situation is more complicated in the presence of matter fields!).

Third, absence of integrations over some set of area tensors $\tau_{\sigma^2}$.
Formally, this looks like the gauge fixing of some tetrad/area components in the
continuous time (3+1) RC or in the continuum general relativity (and indeed reduces to
that gauge fixing in the continuous time limit \cite{Kha2,Kha}). In the fully discrete
theory the reason for fixing $\tau_{\sigma^2}$ is dependence of the curvature matrices
on the t-like triangles on all the rest curvatures via Bianchi identities. It is the
set of those triangles the curvature matrices on which are functions, via Bianchi
identities, of all the rest curvatures. Integrations over $\tau_{\sigma^2}$ might
result in the singularities of the type of $[\delta(R-\bar{R})]^2$. In the situation
when one defines direction of the coordinate called "time", the $t$-like triangles
turn out to be possible choice of the above set. One could keep in mind a regular way
of constructing 4D Regge geometry of the 3D leaves, arbitrary 3D Regge geometries
themselves. For general 4D Regge geometry we adopt the general definition of the set
of "t-like triangles" as the triangles carrying dependent (via Bianchi identities)
curvature matrices. In the present paper we study Regge geometry where"t-like
triangles" will be defined just in this way.

Brief discussion of the above points is given in our work \cite{Kha5}. Expressions
like (\ref{VEV}) present possible exact definition of the formal path integral symbol
in general relativity.

Consider Regge lattice composed of the two identical, up to reflection, 4-simplices
with the vertices 0, 1, 2, 3, 4. We can imagine this complex if consider the two
4-simplices (01234) and $(0^{\prime}1234)$ with common 3-face (1234) and identify the
vertices 0 and $0^{\prime}$. Let all the connections $\Omega$ on the 3-faces act from
(01234) to $(0^{\prime}1234)$, that is, if a 2-face tensor $\pi$ is defined in
(01234), then $\Omega\pi\bar{\Omega}$ is defined in $(0^{\prime}1234)$. Denote a
3-face in the same way as the opposite vertex, $\Omega_i$ $\equiv$ $\Omega_{\sigma^3}$
where $\sigma^3$ = $(\{01234\}\setminus \{i\})$. Here \{\dots\} denote (sub)set, here
of the vertices 0, 1, 2, 3, 4. Denote a 2-face in the same way as the opposite edge,
$v_{(ik)}$ $\equiv$ $v_{\sigma^2}$, $R_{(ik)}$ $\equiv$ $R_{\sigma^2}$ where
$\sigma^2$ = $(\{01234\}\setminus \{ik\})$. It is convenient to define the variables
$v$, $R$ on the ordered pairs of vertices $ik$, $v_{ik}$ = $-v_{ki}$, $R_{ik}$ =
$\bar{R}_{ki}$. (Then $v_{(ik)}$ is a one of the two values, $v_{ik}$ or $v_{ki}$,
$R_{(ik)}$ is $R_{ik}$ or $R_{ki}$). The action takes the form
\begin{equation}                                                                    
S(v,\Omega) = \sum^4_{i<k}{|v_{ik}|\arcsin{v_{ik}\circ R_{ik}\over |v_{ik}|}},
~~~R_{ik} = \bar{\Omega}_i\Omega_k.
\end{equation}

The curvature matrices are related by the Bianchi identities $R_{ik}R_{kl}$ = $R_{il}$
(on the triangles with common edge $(\{01234\}\setminus \{ikl\})$). As independent
curvature matrices we can choose $R_{\alpha}$ $\equiv$ $R_{0\alpha}$ ($\alpha$,
$\beta$, $\gamma$, \dots = 1, 2, 3, 4), that is, the curvature on the 2-faces of the
tetrahedron (1234). Thereby, in accordance with the above definition, the tetrahedron
(1234) can be attributed to the leaf of the foliation (is not t-like) while the other
triangles $(0\alpha\beta)$ can be treated as $t$-like ones. This corresponds to the
naive viewpoint on this Regge lattice as that one corresponding to the evolution in
time of the closed manifold consisting of the two copies of the tetrahedron (1234) to
the point 0. Upon dividing the triangles into the $t$-like ones with the tensors
$\tau_{\alpha\beta}$ $\equiv$ $v_{\alpha\beta}$ entering as parameters and the leaf
ones with the tensors $\pi_{\alpha}$ $\equiv$ $v_{0\alpha}$ the action reads
\begin{equation}                                                                    
S(v,\Omega) = \sum^4_{\alpha = 1}{|\pi_{\alpha}|\arcsin{\pi_{\alpha}\circ
R_{\alpha}\over |\pi_{\alpha}|}} + \sum^4_{\alpha < \beta
}{|\tau_{\alpha\beta}|\arcsin{\tau_{\alpha\beta}\circ (\bar{R}_{\alpha}R_{\beta})\over
|\tau_{\alpha\beta}|}}.
\end{equation}

Due to invariance property of the Haar measure the product of ${\cal
D}\Omega_{\sigma^3}$ over all $\sigma^3$ can be expressed as the product of ${\cal
D}R_{\sigma^2}$ over all not t-like $\sigma^2$ (i.e. $R_{\sigma^2}$ on these
$\sigma^2$ are independent) and ${\cal D}\Omega_{\sigma^3}$ over some set of
$\sigma^3$. In the considered case we can choose
\begin{equation}                                                                    
\prod^4_{i=0}{\cal D}\Omega_i = {\cal D}\Omega_0 \prod^4_{\alpha=1}{\cal D}R_{\alpha}.
\end{equation}

\noindent Integration over ${\cal D}\Omega_0$ is trivial, and our expression for the
vacuum expectation value takes the form
\begin{eqnarray}
\label{VEVmodel}                                                                    
<\Psi (\{\pi\},\{R\})> & = & \int{\Psi (-i\{\pi\}, \{R\})\exp{\left
[-\! \sum^4_{\alpha < \beta}{\tau _{\alpha\beta}\circ
(\bar{R}_{\alpha}R_{\beta})}\right ]}}\nonumber\\
 & & \hspace{-20mm}\times \exp{\left (i
\!\sum^4_{\alpha=1} \pi_{\alpha}\circ R_{\alpha}\right )}\prod^4_{\alpha=1}\d^6
\pi_{\alpha}\prod^4_{\alpha=1}{\cal D}R_{\alpha}.
\end{eqnarray}

Since curvature matrices on classical solutions are not quite physical ones (are more
general than rotations by defect angle around triangles $\sigma^2$), it is natural to
limit ourselves by considering functions to be averaged of purely tensors
$\pi_{\sigma^2}$. More accurately, matrices $R$ whenever appearing among arguments of
a function to be averaged should be substituted by their exact expressions in terms of
$\pi_{\sigma^2}$. Let us perform integration of the measure over curvature matrices.
To this end, we decompose antisymmetric tensor variables into self- and antiselfdual
parts as
\begin{equation}                                                                    
v_{ab} \equiv {1\over 2}\pv_k\pSig^k_{ab} + {1\over 2}\mv_k\mSig^k_{ab}
\end{equation}

\noindent where $\pmSig^k_{ab}$ is basis of (anti)selfdual matrices such that
$i\pmSig^k_{ab}$ obey Pauli matrix algebra. The curvature splits into factors
\begin{equation}
\label{R=exp}                                                                       
\pmR = \exp (\pmbphi\pmbSig) = \cos \pmphi + \pmbSig\pmbn\sin \pmphi,
\end{equation}

\noindent SU(2)-rotations by the angles $\pmphi$ around unit vectors $\pmbn$ =
$\pmbphi/\pmphi$ in the corresponding 3D spaces. In the particular physical case when
$R_{\sigma^2}$ rotates around $\sigma^2$ by the defect angle on $\sigma^2$, both
$\pmphi_{\sigma^2}$ are equal to the half of that defect angle.

The measure splits multiplicatively into the factors in the configuration spaces of
self- and antiselfdual tensor variables,
\begin{eqnarray}
\label{VEVmodel-pm}                                                                 
\d \mu_{\rm area} & = & \d \pmu_{\rm area}\d \mmu_{\rm area}, ~~~ \d \pmmu_{\rm area}
\equiv \d \pmcalN \prod^4_{\alpha=1} \d^3 \pmbpi_{\alpha}, \nonumber\\ \d \pmcalN & =
& \exp \left [-\! \sum^4_{\alpha < \beta} \pmtau_{\alpha\beta}\circ (\overline
{\pmR}_{\alpha}\pmR_{\beta}) - \!\sum^4_{\alpha=1} \pmpi_{\alpha}\circ
\pmR_{\alpha}\right ]\prod^4_{\alpha=1}{\cal D}\pmR_{\alpha}.
\end{eqnarray}

\noindent In what follows, indices $\pm$ on the variables will be suppressed where it
is possible without confusion, and we shall have in view any one of sectors, self- or
antiselfdual one. In the formal limit $\tau_{\alpha\beta}$ = 0 this measure (as well
as the general one (\ref{VEV})) can be factorized into the measures on separate pairs
$\pmpi_{\alpha}$, $\pmR_{\alpha}$. We can apply any such factor to averaging monomials
of the variables and then extend the result to arbitrary functions like
$f(\pi^2_{\alpha})$,
\begin{eqnarray}                                                                    
\int\exp (i\bpi_{\alpha}\bn_{\alpha}\sin\phi_{\alpha}){\sin^2\!\phi_{\alpha}\over
\phi^2_{\alpha}}\d^3\bphi_{\alpha}\d^3\bpi_{\alpha}f(-\pi^2_{\alpha}) & = &
\nonumber\\ & & \hspace{-80mm} \int^{+\infty}
_{-\infty}\cosh^2\!\eta_{\alpha}\d\eta_{\alpha}\int^{\infty}_02\pi\sinh\zeta_{\alpha}
\d\zeta_{\alpha}\int\d^3\bpi_{\alpha}\exp
(-\pi_{\alpha}\cosh\eta_{\alpha}\cosh\zeta_{\alpha})f(\pi^2_{\alpha}).
\end{eqnarray}

\noindent The LHS is the above "$\pi_{\sigma^2}$ $\to$ $-i\pi_{\sigma^2}$"-definition
of the Euclidean integral
\begin{equation}
\label{Eucl-int}                                                                   
\int\exp (-\bpi_{\alpha}\bn_{\alpha}\sin\phi_{\alpha}){\sin^2\!\phi_{\alpha}\over
\phi^2_{\alpha}}\d^3\bphi_{\alpha}\d^3\bpi_{\alpha}f(\pi^2_{\alpha}),
\end{equation}

\noindent while RHS looks as a result of the formal substitution
\begin{equation}
\label{contin}                                                                     
\phi_{\alpha} = \pi/2 + i\eta_{\alpha}, ~~~ \theta_{\alpha} = i\zeta_{\alpha}
\end{equation}

\noindent in (\ref{Eucl-int}), $\theta_{\alpha}$ is the angle between $\bphi_{\alpha}$
and $\bpi_{\alpha}$; $\eta_{\alpha}$, $\zeta_{\alpha}$ are real. Let us integrate over
contours (\ref{contin}) in the case $\tau_{\alpha\beta}$ $\neq$ 0 of interest. The
formula (\ref{VEVmodel-pm}) gives
\begin{eqnarray}
\label{VEVmodel-analyt}                                                            
\pmcalN & \equiv & \int \d \pmcalN = \int \exp\left [
-\sum^4_{\alpha=1}\pi_{\alpha}z_{\alpha}\cosh \eta_{\alpha} + \sum^4_{\alpha <
\beta}i\btau_{\alpha\beta}\cdot (\bn_{\alpha}\cosh \eta_{\alpha} \sinh \eta_{\beta} -
\right. \nonumber\\ & & \hspace{-25mm} \left. - \bn_{\beta}\cosh \eta_{\beta} \sinh
\eta_{\alpha}) + \sum^4_{\alpha < \beta} \btau_{\alpha\beta} \cdot (\bn_{\alpha}
\times \bn_{\beta}) \cosh \eta_{\alpha} \cosh \eta_{\beta} \right ] \prod^4_{\alpha=1}
\cosh^2\eta_{\alpha} \d \eta_{\alpha} \d z_{\alpha} \d \chi_{\alpha}
\end{eqnarray}

\noindent where $z_{\alpha}$ = $\cosh \zeta_{\alpha}$, $\chi_{\alpha}$ is polar angle
of $\bphi_{\alpha}$ w. r. t. $\bpi_{\alpha}$. Let us write the exponent in
(\ref{VEVmodel-analyt}) in the form
$\exp(-\sum^4_{\alpha=1}\pi_{\alpha}z_{\alpha}\cosh\chi_{\alpha} + f)$ and integrate
over $\d z_{\alpha}$ by parts, first integrating
$\exp(-\sum^4_{\alpha=1}\pi_{\alpha}z_{\alpha}\cosh\chi_{\alpha})$ and differentiating
$\exp f$. We arrive at
\begin{eqnarray}
\label{int-dz}                                                                     
\int^{\infty}_1\exp\left (-\sum^4_{\alpha=1}\pi_{\alpha}z_{\alpha}\cosh\eta_{\alpha} +
f \right )\d z_{\alpha} & = & \nonumber\\ & & \hspace{-40mm} {1\over \pi_{\alpha}\cosh
\eta_{\alpha}}\exp \left. \left
(-\sum^4_{\alpha=1}\pi_{\alpha}z_{\alpha}\cosh\eta_{\alpha} + f \right )\right
|^{z_{\alpha} = 1}_{z_{\alpha} = \infty} + O(\tau)
\end{eqnarray}

\noindent The function $f$ = $O(\tau)$. At $\tau_{\alpha\beta}$ $\to$ 0 the value in
the RHS of (\ref{int-dz}) taken at $z_{\alpha}$ = $\infty$ is well-defined and is zero
in the range of changing the arguments $\pi_{\alpha}$ tending to the whole range of
$\pi_{\alpha}$. The function $f$ is up to $O(z^{-1}_{\alpha})$ linear in $z_{\alpha}$,
and situation looks like calculation of the integral
\begin{equation}                                                                   
\int^{\infty}_1 \exp (-\lambda z)\d z = {\exp (-\lambda)\over \lambda},
\end{equation}

\noindent at $\lambda$ $<$ 0 it is considered as analytical continuation from the
region $\lambda$ $>$ 0. In the case of the formula (\ref{int-dz}) this means that the
value in the RHS taken at $z_{\alpha}$ = $\infty$ should be treated as zero.

Thus, the measure splits into the terms vanishing in the formal limit
$\btau_{\alpha\beta}$ = 0 and an expression taken at $z_{\alpha}$ = 1. The values
$z_{\alpha}$ = 1 mean that $R_{\alpha}$ rotate around corresponding triangles, i.e.
have exact physical sense of curvature matrices. Some disadvantage of our highly
nonperturbative model having only one 3D leaf (another one is contracted to the point
0) is that some components $|\btau_{\alpha\beta}|$ cannot be small on physical
hypersurface (when tetrahedrons are closed) if $\pi_{\alpha}$ grow. However, we
concentrate on studying just the $z_{\alpha}$ = 1 term in the RHS of (\ref{int-dz})
having in view its relevance to the more usual Regge structures where the condition
$|\btau_{\alpha\beta}|$ $\ll$ 1 is quite accessible by using closely located and
similar in their geometry 3D leaves.

At $z_{\alpha}$ = 1 integrations over polar angles $\chi_{\alpha}$ are trivial, and we
find (up to normalization, as usual)
\begin{eqnarray}
\label{VEVmodel-z=1}                                                               
\pmcalN & = & \int \exp\left [ -\sum^4_{\alpha=1}\pi_{\alpha}\cosh \eta_{\alpha}
 + \sum^4_{\alpha < \beta} \btau_{\alpha\beta} \cdot (\bn_{\alpha}
\times \bn_{\beta}) \cosh \eta_{\alpha} \cosh \eta_{\beta} \right ] \nonumber\\ & &
\hspace{-20mm}\times \cos\left [\sum^4_{\alpha < \beta}\btau_{\alpha\beta}\cdot
(\bn_{\alpha}\cosh \eta_{\alpha} \sinh \eta_{\beta} - \bn_{\beta}\cosh \eta_{\beta}
\sinh \eta_{\alpha}) \right ] \prod^4_{\alpha=1} {1\over \pi_{\alpha}} \cosh
\eta_{\alpha} \d \eta_{\alpha}
\end{eqnarray}

\noindent where from now on $\bn_{\alpha}$ = $\bpi_{\alpha}/\pi_{\alpha}$. Dangerous
for convergence is the bilinear in $\cosh \eta_{\alpha}$ term in the exponential. What
is its sign? It is arbitrary until $\btau_{\alpha\beta}$ are specified, namely, until
area tensors are chosen to correspond to certain edge vectors. In this case
$\tau_{\alpha\beta}$ are parameterized as follows,
\begin{equation}
\label{tau=pi,pi}                                                                  
\tau_{\alpha\beta} = u_{\beta}\pi_{\alpha} - u_{\alpha}\pi_{\beta} + \lambda
[\pi_{\beta}, \pi_{\alpha}]
\end{equation}

\noindent where
\begin{equation}                                                                   
\sum^4_{\alpha=1} u_{\alpha} = 1, ~~~ \sum^4_{\alpha=1} \pi_{\alpha} = 0.
\end{equation}

\noindent This anzats solves the conditions on area tensors of the type
\begin{equation}                                                                   
v_{\sigma^2_1} \ast v_{\sigma^2_2} \equiv {1\over 4} \epsilon_{abcd}
v^{ab}_{\sigma^2_1} v^{cd}_{\sigma^2_2} = 0
\end{equation}

\noindent for any two 2-simplices $\sigma^2_1$, $\sigma^2_2$ sharing a common edge,
and
\begin{equation}                                                                   
v_{\sigma^2_1} \ast v_{\sigma^2_2} = \pm v_{\sigma^2_3} \ast v_{\sigma^2_4}
\end{equation}

\noindent for any two pairs of 2-simplices of the same 4-simplex such that
intersections in these pairs, $\sigma^2_1 \cap \sigma^2_2$ and $\sigma^2_3 \cap
\sigma^2_4$, are 0-simplices (vertices). In the present model notations
\begin{eqnarray}
\label{pi*tau}                                                                     
\pi_{\alpha} \ast \pi_{\beta} = 0, ~~~ \tau_{\alpha\beta} \ast \pi_{\alpha} = 0, ~~~
\tau_{12} \ast \pi_3 = \tau_{23} \ast \pi_1 = \tau_{31} \ast \pi_2.
\end{eqnarray}

\noindent Irreducible set of 20 constraints follows if $\alpha$, $\beta$ run over 1,
2, 3 here. Self-antiselfdual splitting recasts (\ref{tau=pi,pi}) to vector form,
\begin{equation}                                                                   
\btau_{\alpha\beta} = u_{\beta}\bpi_{\alpha} - u_{\alpha}\bpi_{\beta} + \lambda
\bpi_{\beta} \times \bpi_{\alpha}
\end{equation}

\noindent so that the sign of the last term in the exponential
\begin{equation}                                                                   
\exp \left [ -\sum^4_{\alpha=1} \pi_{\alpha} \cosh \eta_{\alpha} - \lambda
\sum^4_{\alpha < \beta} (\bn_{\alpha} \times \bn_{\beta})^2 \pi_{\alpha} \pi_{\beta}
\cosh \eta_{\alpha} \cosh \eta_{\beta} \right ]
\end{equation}

\noindent is defined by the sign of $\lambda$. Integral over $\d \eta_{\alpha}$ is
absolutely convergent at $\lambda$ $\geq$ 0 and reminds in this respect
\begin{equation}                                                                   
\int^{\infty}_1 \exp (-x - \lambda x^2) \d x,
\end{equation}

\noindent nonanalytical at $\lambda$ = 0 and complex if continued to $\lambda$ $<$ 0.
In our case $\lambda$ = 0 means zero volume of the 4-simplex, i.e. degenerate metric.

Thus, the measure (\ref{VEVmodel-z=1}) being reduced to the physical hypersurface
(ordinary RC) and integrated over connections can be well-defined on condition that we
treat 4-simplices as nondegenerate and preserving orientation. The resulting measure
possesses exponential (w. r. t. areas) cutoff. Moreover, on physical hypersurface the
resulting full measure proportional to $\pcalN\mcalN$ is {\it positive}. This is
simply because the factors $\pcalN$ and $\mcalN$ are equal to the same function of the
arguments $\pbpi_{\alpha}\cdot\!\pbpi_{\beta}$, $\pbtau_{\alpha\beta}\cdot
\!\pbpi_{\alpha}$ and $\mbpi_{\alpha}\cdot\!\mbpi_{\beta}$,
$\mbtau_{\alpha\beta}\cdot\!\mbpi_{\alpha}$, respectively. On physical hypersurface
the "+" and "$-$" counterparts from this set are equal to the same value due to
(\ref{pi*tau}) since
\begin{equation}                                                                   
v_1 \ast v_2 = {1 \over 2} \pbv_1\cdot\!\pbv_2 - {1 \over 2} \mbv_1\cdot\! \mbv_2.
\end{equation}

\noindent Therefore $\d \mu_{\rm area}$ turns out to be proportional to the square of
a real value.

We have illustrated by the RC model of a simple structure that upon going to the
complex integration contours in the space of matrices $\Omega$ the quantum measure
could be well-defined. Matrices $\Omega_{\sigma^3}$ in our formalism are more general
values than physical connections since the curvature matrices $R_{\sigma^2}$ composed
of $\Omega_{\sigma^3}$ being classical solution of the equations of motion do not
necessarily rotate around the triangle $\sigma^2$ (analog of the continuum curvature
with torsion). Therefore the curvature matrices entering any physical quantity to be
averaged should be considered not as independent ones but as already expressed in
terms of area tensors. Then really independent matrices $\Omega_{\sigma^3}$ could be
integrated out from the very beginning. The resulting measure on physical hypersurface
is positive and possesses exponential cutoff at large areas.

The integration contours in the complex connection space pass through independent
matrices $R_{\sigma^2}$ corresponding to rotations by the angles with real part $\pi
/2$, while the axes of these rotations form the purely imaginary angles with
$\sigma^2$ area vectors. These contours are strictly separated from trivial curvature
matrices and in this sense the analytical continuation could be called nonperturbative
one. We have found that up to $O(\tau)$ the integrated over connections quantum
measure is defined by its value at the points in the superspace of connections
corresponding to independent matrices $R_{\sigma^2}$ rotating {\it around} triangles
$\sigma^2$, i.e. around area $\sigma^2$ vectors in the self-antiselfdual
representation.

The main problem is whether the matrices $R_{\sigma^2}$ on the t-like triangles
expressed by Bianchi identities in terms of the above complex nonperturbative
rotations could provide the contributions $\tau_{\sigma^2} \circ R_{\sigma^2}$ in the
exponential ensuring convergence of the integral over connections. Here we have given
affirmative answer for the simplest Regge structure and the more general structures
remain to be studied.

\bigskip

The present work was supported in part by the Russian Foundation for Basic Research
through Grant No. 05-02-16627-a.

\end{document}